\documentclass[a4paper,fleqn,usenatbib]{mnras}

\usepackage{newtxtext,newtxmath}

\usepackage[T1]{fontenc}
\usepackage{ae,aecompl}
\usepackage{bm}

\usepackage{graphicx}	
\usepackage{amsmath}	
\usepackage{amssymb}	
\usepackage{chemmacros}
\usepackage{color}
\usepackage[version=4]{mhchem}
\usepackage[flushleft]{threeparttable}

\title[Lightning chemistry on Earth]{Lightning Chemistry on Earth-like Exoplanets}

\author[Ardaseva et al.]{
Aleksandra Ardaseva,$^{1}$
Paul B. Rimmer,$^{1}$\thanks{E-mail: pr33@st-andrews.ac.uk}
Ingo Waldmann,$^{2}$
Marco Rocchetto,$^{2}$ 
\and Sergei N. Yurchenko,$^{2}$  
Christiane Helling,$^{1}$ 
Jonathan Tennyson$^{2}$
\\
$^{1}$Centre for Exoplanet Science, SUPA, School of Physics and Astronomy, University of St Andrews, North Haugh, St Andrews, KY16 9SS, UK\\
$^{2}$Department of Physics and Astronomy, University College London, London, WC1E 6BT, UK
}

\date{Accepted XXX. Received YYY; in original form ZZZ}

\pubyear{2017}

\begin{document}
\label{firstpage}
\pagerange{\pageref{firstpage}--\pageref{lastpage}}
\maketitle

\begin{abstract}
{We present a model for lightning shock induced chemistry that can be
  applied to atmospheres of arbitrary H/C/N/O chemistry, hence for
  extrasolar planets and brown dwarfs. The model couples hydrodynamics
  and the STAND2015 kinetic gas-phase chemistry. For an exoplanet analogue 
  to the contemporary Earth,
   our model predicts \ce{NO} and \ce{NO_2} yields in agreement with
   observation. We predict height-dependent mixing ratios during a storm 
   soon after a lightning shock of \ce{NO} $\approx 10^{-3}$ at 40 km and \ce{NO_2} $\approx
   10^{-4}$ below 40 km, with \ce{O3} reduced to trace quantities
   ($\ll 10^{-10}$). For an Earth-like exoplanet with a
   \ce{CO_2}/\ce{N_2} dominated atmosphere and with an extremely intense lightning storm over its entire surface, we predict significant changes in the amount of
   \ce{NO}, \ce{NO_2}, \ce{O_3}, \ce{H_2O}, \ce{H_2}, and predict
   significant abundance of \ce{C_2N}. We find that, for the Early Earth, \ce{O_2} is formed in large quantities by lightning but is rapidly processed by the
   photochemistry, consistent with previous work on lightning. The
   effect of persistent global lightning storms are predicted to be
   significant, primarily due to \ce{NO_2}, with the largest
   spectral features present at $\sim 3.4$ $\mu$m and $\sim 6.2$ $\mu$m. The
   features within the transmission spectrum are on the order of 1 ppm and
   therefore are not
   likely detectable with JWST. 
    Depending on its spectral properties, \ce{C_2N} could be a key
    tracer for lightning on Earth-like exoplanets with a
    \ce{N_2}/\ce{CO_2} bulk atmosphere, unless destroyed by yet
    unknown chemical reactions.
}
\end{abstract}





\section{Introduction} \label{intro}

A large number of exoplanets have been discovered over the
  last few years\footnote{\url{http://exoplanet.eu/}}. They
differ by the location, characteristics of the host star, and both
chemical and physical compositions. Recently, candidate rocky planets
within the ``liquid water'' habitable zone of their star have been discovered: the seven TRAPPIST planets \citep{Gillon2016,Gillon2017}; and Proxima b
\citep{Anglada2016}. Transmission spectra of TRAPPIST-1b and TRAPPIST-1c hint at a cloudy atmosphere obscuring spectral
signatures \citep{deWit2016}.  The discovery of these planets has
further propelled already significant interest into the composition of
atmospheres and clouds on potentially habitable exoplanets
\citep[e.g][]{Kreidberg2016}. \citet{Kane2016} have made the first
attempt at compiling a catalogue of potentially habitable exoplanets,
using various definitions of the liquid water habitable zone as the
identifying criteria.

Clouds, and physical processes related to clouds, are of great interest for determining how probable it would be for life to have arisen on rocky exoplanets, by stabilizing the temperature and effectively expanding the habitable zone \citep{Yang2013} and by introducing the possibility for lightning discharges, which may generate prebiotic chemistry \citep{Miller1953a}. An increasing body of evidence shows that the presence of clouds in exoplanet atmospheres is ubiquitous \citep[e.g.][]{Bean2010,Sing2011,Wordsworth2011,Radigan2012}. These clouds are comprised of particles made of a mix of materials at a rich variety and for which there are often no clear analogues to be found within our solar system. 

The structure and composition of the clouds change dependent on the local
thermodynamic conditions and the availability, or lack, of nucleating sites such as ocean spray, volcanic ash, and sand. Also the cloud particle size distribution changes over the extent of the atmosphere and over time. The cloud dynamics, in conjunction with charging processes, can result in significant electric fields spanning large distances. This is because cloud particles carry an excess of positive or negative charge over a great distance, resulting in a large-scale charge separation. The electric field may initiate a discharge, such as lightning, in order to restore that balance \citep{Beasley1982}. 

At present, there is only one definitive example of a habitable planet, the Earth, and therefore our present investigation into exoplanetary lightning will focus on Earth analogues. Understanding lightning on these planets, compared to lightning on Earth, is potentially important both for investigating habitability beyond the question of surface liquid water, and for gaining insight into the physical processes on rocky exoplanets, such as exoplanetary global electric circuits \citep{Helling2016}.

Earth-like exoplanets, similarly sized rocky planets with a Sun-like
host star, have an occurrence rate of 0.51 planet per star estimated
from statistics on the available sample of exoplanets
\citep{Dressing2013}. The spectral features of Earth-like explanets have already been extensively modelled for a diversity of UV fields by \citet{Rugheimer2013}. The varying amount of water on the surface is
predicted to have a considerable effect on the rate of lightning. We expect dry, rocky planets to have lightning flash densities
equal to $17.0-28.9\,\textup{flashes}\,\textup{km}^{-2}\,\textup{yr}^{-1}$, whereas Earth-sized planets containing water on their surface would show smaller frequency of only $0.3-0.6\,\textup{flashes}\,\textup{km}^{-2}\,\textup{yr}^{-1}$ \citep{Hodosan2016}. 

One of the most detailed observational studies of lightning on Earth
was carried by \cite{Orville1968a, Orville1968b, Orville1968c}. Orville performed a time-resolved spectroscopy with
the resolution of $5\,\mu\textup{s}$ on multiple lightning
flashes. Using NII emission lines, Orville approximated the peak value
of the temperature to lie within the range of
$28000-31000\,\textup{K}$. This value is obtained from 7 flash spectra
and the peak temperature is widely accepted to be $T_{\rm gas}=30000\,\textup{K}$. The number density inside the
lightning channel is estimated from the H$\alpha$ broadening, assuming
the broadening is caused by the Stark effect only. The spectrum of
only one lightning flash showed this feature, therefore it is
difficult to determine the uncertainty in the number density; the
temperature is better constrained. In Orville's model, the peak
pressure is approximated to equal $P_{\rm in}=8\,\textup{atm}$, when
the pressure of ambient medium is $P_{\rm gas}=1\,\textup{atm}$. The pressure is determined experimentally from the experimentally measured
equation of state of air at temperatures up to $24000$ K \citep{Gilmore1955}.

High temperatures in the lightning channel are very favourable for the dissociation of molecular nitrogen \ce{N_2} -- a very stable molecule with the dissociation energy being $9.756\,\textup{eV}$ \citep{Frost1956}. The separated nitrogen atoms then participate in neutral Zel'dovich reactions \ref{eq:zeldov1} and \ref{eq:zeldov2} to form nitric oxide \citep{Zeldovich2002}.
\begin{gather}  \label{eq:zeldov2}
\textup{N}+\textup{O}_2 \rightarrow \textup{NO}+\textup{N} \\
\label{eq:zeldov1}
\textup{N}_2 + \textup{O} \rightarrow \textup{NO} + \textup{O} 
\end{gather}
\cite{Borucki1984} conclude that approximately $10^{10}\,\textup{kg}$ of both \ce{NO} and \ce{NO_2} is produced in the atmosphere of Earth per year as a consequence of thunderstorms. This makes nitric oxide a signature molecule of lightning on present-day Earth. \citet{Price1997} observed and characterised the chemical impact of lightning on the atmosphere of the contemporary Earth. This work showed that the effect of lightning on \ce{NO} and \ce{NO_2} is dwarfed by the anthropogenic sources of these molecules.

The ability to dissociate \ce{N_2} also provides a potential route for the formation of complex molecules and amino acids, as shown in the Miller-Urey experiment as long as it occurs in favourable chemical environment. Experimentally, the lightning is investigated using laser-induced plasma (LIP) \citep{Jebens1992, Navarro-Gonzalez2001}. This approach allows to reach temperatures inside the channel up to $\approx10^4$ K and has provided insight on how to best link observations of lightning-induced chemistry to theoretical models.

A detailed model of the chemical impact of lightning shocks for atmospheres of a range of compositions will be needed in order to find what effect various flash rates would have on the global chemistry for the diverse set of observed exoplanets. In addition, a coupled hydrodynamic shock model of lightning and chemical shock model will be useful for studies of atmospheric chemistry during lightning storms and of the effect of lightning on chemistry for the Early Earth.

In order to aid in this investigation, we present a lightning model, in which we take existing hydrodynamic and chemical kinetic models of lightning shock induced chemistry, and couple them in order to predict the chemical effects of lightning within atmospheres of arbitrary H/C/N/O chemistry. Our's is the only atmospheric model to account for lightning beyond adding chemical lightning yields as source terms within the atmosphere. Given our focus in this work, this model is here applied specifically for the Contemporary and Early Earth, and makes new predictions for lightning on both the Contemporary and Early Earth. We determine the impact of an intense global lightning storm on the transmission and emission spectrum of Earth-like exoplanets and then discuss the possibility of observing spectral signatures of lightning in this extreme case.

We start by describing the computational model and initial conditions in Section \ref{methods}. The setup of hydrodynamical shock model and a follow-up comparison with Orville's data is discussed in \ref{hydrodyn}. The chemical kinetics network STAND is described in \ref{stand}. We then proceed in \ref{modern} with an overview and a discussion of the resulting impact of lightning onto the Contemporary Earth atmosphere. The section \ref{Early} includes the analysis of lightning on Early Earth atmosphere. The hypothetical spectra of Earth-like planets with strong lightning activity is presented in section \ref{spectra}.

\section{Approach} \label{methods}
Here we lay out the methodology for our coupled hydrodynamic chemical kinetics model of the lightning discharge. We use the \textsc{ATHENA} MHD code \citep{Stone2008} to develop the 2D lightning shock model. ATHENA implements algorithms that allow the use of static and adaptive mesh refinement which solves the conservation of mass, momentum, and energy through the grid (see Appendix \ref{appendix:app1}). The code has been extensively tested, including for the shock tube in 1D, Rayleigh-Taylor instabilities in 2D and 3D \citep{Stone2008}. The initial conditions of the lightning are taken from the observations carried by \citet{Orville1968b}. To predict the chemistry, we use the \textsc{STAND2015} chemical network constructed for lightning shock chemistry along with the \textsc{ARGO} photochemistry/diffusion solver \citep{Rimmer2016}, a Lagrangian solver that has recently been validated against the standard Eulerian solvers \citep{Tsai2016}. The chemical network solves for H/C/N/O chemistry and has been successfully benchmarked against both Contemporary and Early Earth models \citep{Rimmer2016}. We apply these approaches to specific temperature profiles and bulk atmospheric compositions appropriate for the Early and Contemporary Earth.

\subsection{Hydrodynamical shock modelling} \label{hydrodyn}

We use \textsc{ATHENA} to set up a 2D hydrodynamical simulation of the shock waves propagation during the lightning. We explored a 1D and a 2D model setup which allowed us to demonstrate the stability of the hydrodynamic solution of our shock wave model.  The 2D setup further allows for a first visualisation of the mainly 2D geometry of the lighting channel.

The computational grid contains 400x400 cells, each corresponding to 1 $\textup{cm}^2$. The initial width of lightning channel is set to 1 grid cell. Two shock waves then propagate parallel in opposite directions, assuming open boundary conditions. For every point in time and space, \textsc{ATHENA} solves the equation of hydrodynamics (Appendix \ref{appendix:app1}). We assume that the cooling of the lightning channel occurs mainly due to radiative processes. Thus, we incorporate a radiative cooling function, discussed in Appendix \ref{appendix:app2}.

\begin{figure}
  \centering
    \includegraphics[width=\columnwidth]{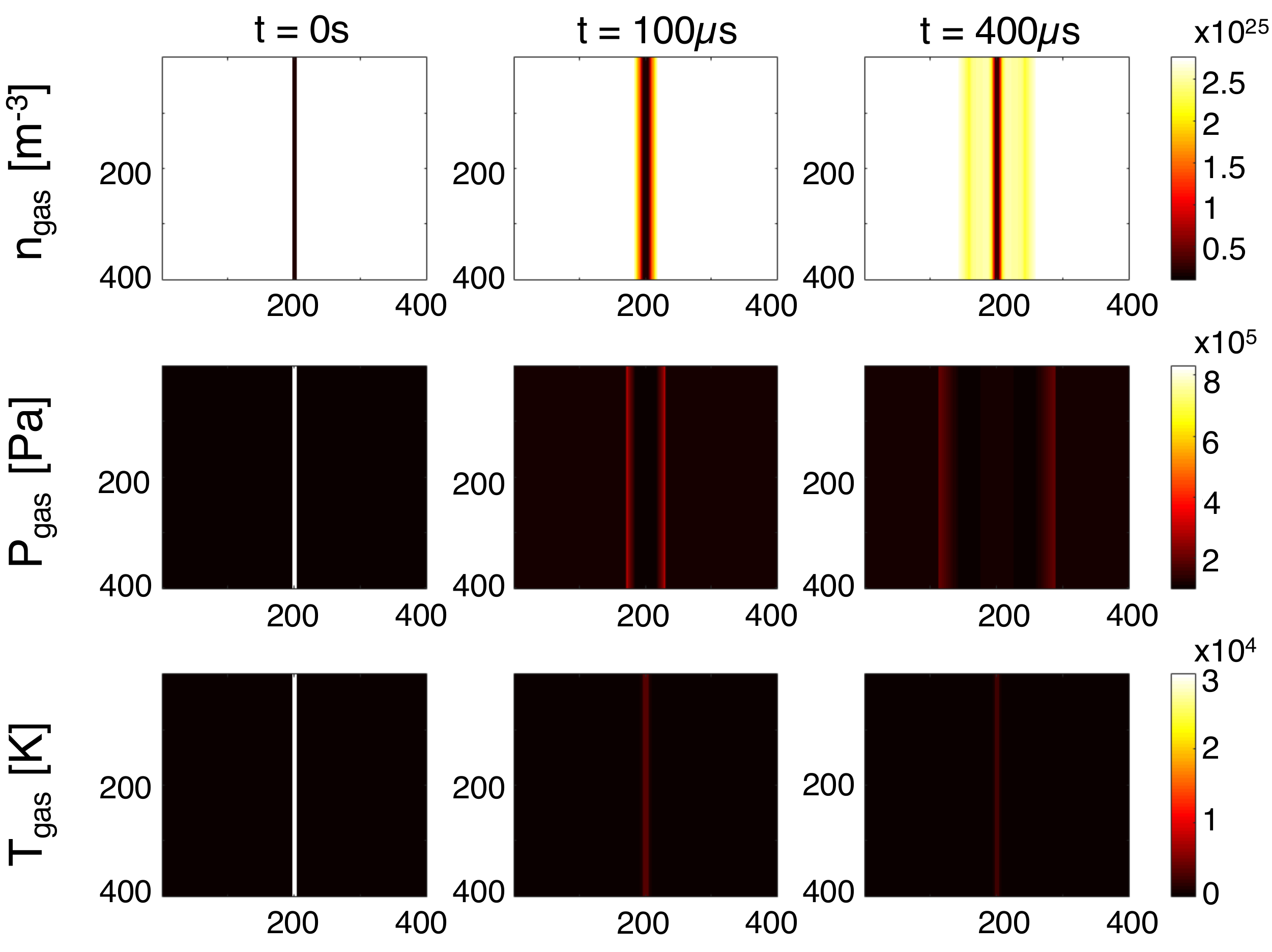}
    \caption{2D snapshots of lightning shock wave propagation at $0\, \textup{s}$, $100\,\mu \textup{s}$, and $400\,\mu \textup{s}$. Spatial axis have units of $\textup{cm}$. First row shows changes in the gas number density, second -- in gas pressure, and third -- in gas temperature.}
\label{fig:hydro}
\end{figure}

We assume the ideal gas with $\gamma=1.44$. The initial conditions inside the lightning channel are set to $T_{in}=30000\,\textup{K}$ and $P_{in}=8\textup{P}_{\rm gas}$ according to \cite{Orville1968b}. The lightning is initiated at 0 km altitude where the physical conditions of the air are the following: $P_{gas}=103900\,\textup{Pa}$ and $T_{gas}=272.1\,\textup{K}$. The atmosphere is composed mostly of oxygen and nitrogen with traces of other elements, with the molar mass $M = 28.97\,\textup{mol}\,\textup{g}^{-1}$. Figure \ref{fig:hydro} shows the changes in $n_{gas}$, $P_{gas}$, $T_{gas}$ during the first 400 $\mu\textup{s}$.

\begin{figure}
  \centering
    \includegraphics[width=\columnwidth]{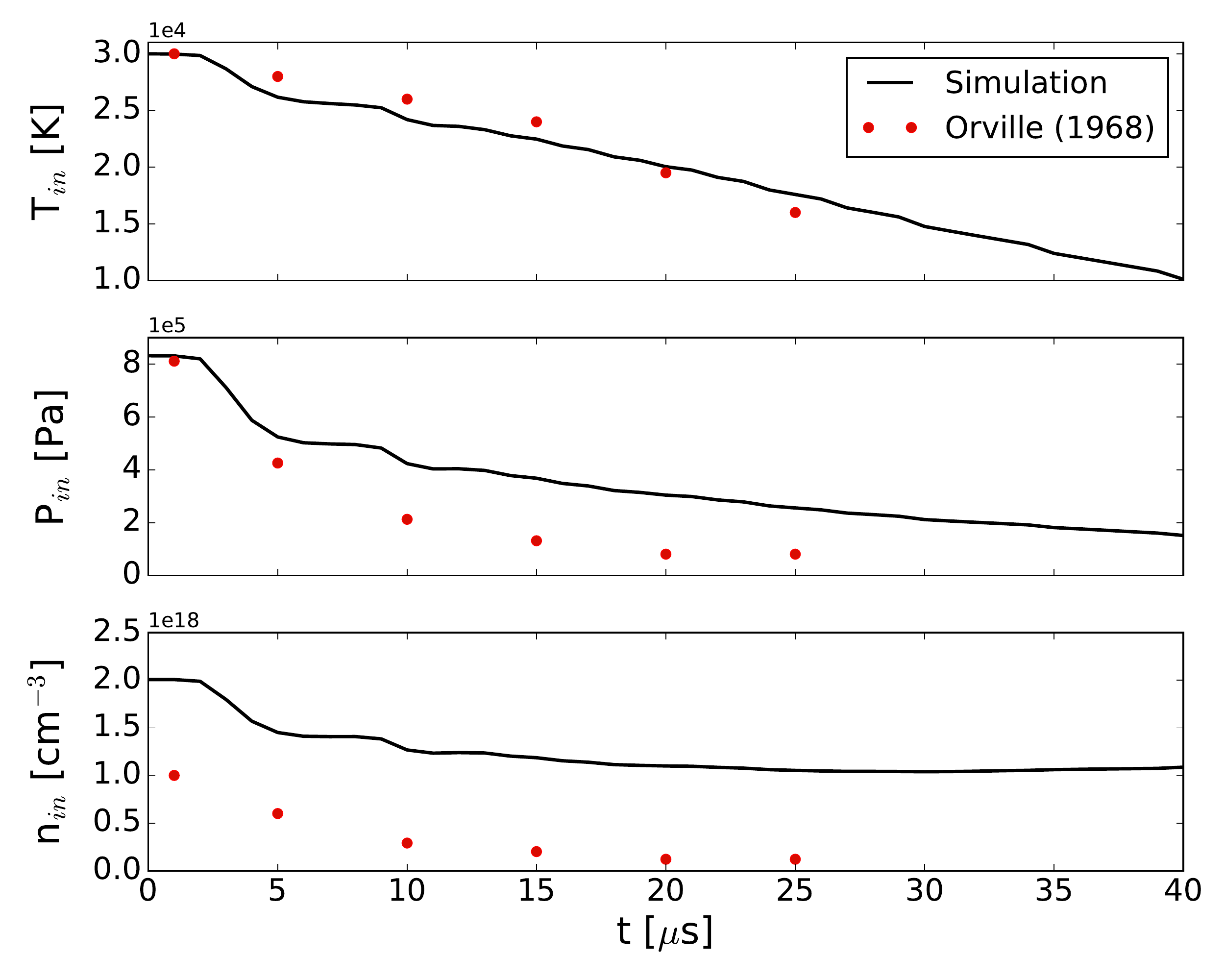}
    \caption{Temperature, $T_{in}$, pressure, $P_{in}$, and number density, $n_{in}$, profiles of the lightning in the first 40 $\mu\textup{s}$. Blue line -- simulation results, red dots -- Orville's data. The deviation in number density values is a result of differences between our simulation and the estimate of \citet{Orville1968c}.}
\label{fig:orville}
\end{figure}

The $P_{in}$, $n_{in}$, $T_{in}$ values are extracted from the centre of the initial discharge channel, i.e. from 200x200th grid cell, and are presented in the Figure \ref{fig:orville}. The simulation demonstrates a good fit with observational temperature data. Orville's temperature values are believed to be precise and reliable since are obtained from NII emission line of multiple flash spectra. The number density curve obtained in the simulations show the difference with Orville's by a factor of 2. Orville estimated the number density from H-alpha broadening due to Stark effect. Only one flash produced detectable spectral features, thus large errors are expected. The differences between simulation and observations is also observed in pressure curve. This is because our simulation explicitly respects the ideal gas law, whereas Orville estimates the pressure independently. The pressure values are taken from prior measurements of air at temperatures below $24000$ K \citep{Gilmore1955}, which is much lower than the estimated temperature inside the discharge channel.

All numerically obtained values of $P$, $n$, $T$ never go to unreasonably low or high values, and the physical conditions of the medium return to pre-shocked values after 0.8 s. The simulation values of $P$, $n$, and $T$ agree with ideal gas law, thus are considered and the $P_{in}(t)$, $n_{in}(t)$, $T_{in}(t)$ output is used as an input for chemical kinetics network. 

\subsection{Chemical kinetics network} \label{stand}

The lightning shock model is coupled to a 1D photochemistry-diffusion code ARGO \citep{Rimmer2016}, which solves the continuity equations for vertical atmospheric chemistry:
\begin{equation}
    \dfrac{\partial n_i}{\partial t} = P_i - L_i - \dfrac{\partial \Phi_i}{\partial z},
\end{equation}
where $n_i(t,z)$ [cm$^{-3}$] is the number density of species $i$ and $i = 1, ..., N_s$, and $N_s$ is the total number of species. $P_i$ is the rate of production and $L_i$ is the rate of loss, both with units cm$^{-3}$ s$^{-1}$. The vertical change in flux, $\Phi_i$, represents both Eddy and molecular diffusion. Except at one height in the atmosphere, this equation is solved precisely as described in \citet{Rimmer2016}, where each height has a constant gas temperature, $T$ [K], and pressure, $p$ [bar]. A parcel of gas is followed as it moves through the atmosphere, and chemistry is tracked for this parcel. The chemical network used here is the \textsc{Stand2016} network \citep{Rimmer2016}, except for R227/228:
\begin{equation*}
\ce{CH_3} + \ce{OH} + \ce{M} \rightarrow \ce{CH_3OH} + \ce{M}
\end{equation*}
for which we replace the rate coefficient used in \citet{Rimmer2016} with the rate coefficient from \citet{Jasper2007}, following \citet{Tsai2016}.

Lightning shock chemistry is initiated deep in the atmosphere, where $p \sim 0.1 - 1$ bar. As discussed in Section 2.1, at a set height, $z_l = 0$ km, the lightning shock is initiated. Right when the parcel achieves this height, the lightning shock initiates. Pressure and temperature are determined in the manner described above in Section \ref{hydrodyn}. The rate of change of these parameters, the magnitude of the temperature and the high pressures are such that any chemical timescale, or even the timescale for the lightning shock itself, is much shorter than the dynamical timescales. Although molecular diffusion is considered alongside the shock chemistry in the code, effectively, for the duration of the lightning shock, $\Phi_i = 0$.

Lightning achieves temperatures on the order of $30000$ K at the
beginning of our shock, and these temperatures will effectively
dissociate and ionize the gas. To account for this without overly
taxing the integrator, we set the initial conditions for the parcel at
$z_l = 0$ km such that all species are fully dissociated and
ionized. The elemental abundances are maintained, but entirely in the form of singly ionized cations. This initial condition deviates from our assumptions for the cooling rate (see
Appendix \ref{appendix:app2}). For every cation,
an electron, $e^-$, is introduced to preserve charge balance. This
means that our assumed initial conditions are such that the degree of ionization, $f_e = 1$. This may seem artificial, but is not unreasonable if one considers  that every collision in a 30000 K gas will be dissociating
and/or ionizing. The barriers for dissociation and ionization are on the same order as the temperature, so the barrier will only slow things by at
most a factor of $\sim \exp(20$ eV $/kT) \sim 10^3$, for the highest
ionization potentials. The time-scale for complete ionization in a
30000 K, 8 bar gas is therefore on the order of:
\begin{equation*}
\dfrac{e^{-I/kT}}{\sigma v n_{\rm gas}} \approx 10^{-9} \; {\rm s}.
\end{equation*}
 This is vastly shorter than the time resolution in which we consider our lightning shock. After the temperature falls to $\sim 10000$ K, however, chemical timescales extend to the length of seconds, and at much lower temperatures, potentially to days or years. In this manner, lightning chemistry can linger for an extended period of time and can affect the entire atmospheric chemistry above which lightning has recently been initiated.

The competition between the rate of lightning events and the chemical time-scales, largely set by the pressures and the energetic barriers for destroying species generated in the lightning shock at quantities far outside equilibrium, will determine the global mixing ratios of lightning species. This sort of analysis would require detailed lightning statistics, of the form of, e.g. \citet{Hodosan2016}. As a first step, we do not consider this detailed statistics, but rather assume every parcel of gas at the exoplanetary surface receives a lightning shock.

A parcel experiences the shock at $z = 0$ km, and remains at this height for a time set by the dynamical timescale, $t_d$ [s]. The dynamical timescale is determined from the Eddy Diffusion coefficient, $K_{zz}$ [cm$^2$ s$^{-1}$] and the difference between the heights at which the constant temperatures and pressures are set, $\Delta h$ [km], as follows:
\begin{equation}
t_d = \dfrac{(\Delta h)^2}{K_{zz}}
\end{equation}
At $z = 0$, $K_{zz} = 10^5$ cm$^{2}$ s$^{-1}$ and $\Delta h = 2$ km, so $t_d \approx 4.6$ days. Therefore, the parcel receives the shock, remains at $z = 0$ km for 4.6 days, and then moves up to a new height. Right before the parcel is moved, its chemistry is recorded, and the entire region at $z = 0$ km is treated as having the final chemistry of this parcel. This is effectively setting the time between each lightning event that the gas at $z = 0$ km experiences equal to 4.6 days. From this timescale and the average energy of a lightning flash, we can work out the effective lightning density assumed for our model.

The flash density, $\rho_{\rm fl}$ [flashes km$^{-2}$ h$^{-1}$] will be proportional to the number density of the gas at the shock $n_{\rm in}$ [cm$^{-3}$] multiplied by the energy added to each particle by the shock, which by the equipartition theorem we will set to $3/2 k_B \Delta T \approx 
3/2 k_B T_{\rm in}$. Treating this as an ideal gas:
\begin{align}
\rho_{\rm fl} &= \Delta h \, \dfrac{P_{\rm in}}{k_B T_{\rm in}} \, \dfrac{\frac{3}{2}k_B T_{\rm in}}{E_{\rm fl}} \, \dfrac{K_{zz}}{(\Delta h)^2},\\
&= \dfrac{3 P_{\rm in} K_{zz}}{2 \Delta h E_{\rm fl}},
\end{align}
where we take $E_{\rm fl} = 4 \times 10^{8}$ J as the average energy of a lightning flash \citep{Borucki1984}. This provides a value of $\rho_{\rm fl} \approx 5.4 \times 10^4$ flashes km$^{-2}$ h$^{-1}$. By comparison, the highest flash density observed on Earth is $\sim0.1$ flashes km$^{-2}$ h$^{-1}$ produced in thunderstorms in Florida and other areas in the United States \citep{Huffines1999}.
We consider these intense lightning flash densities to be a practical
upper limit for estimating the chemical impact of lightning on Earth-like exoplanets with Earth-like flash energies.

\subsection{Initial conditions}

\begin{figure}
  \centering
    \includegraphics[width=\columnwidth]{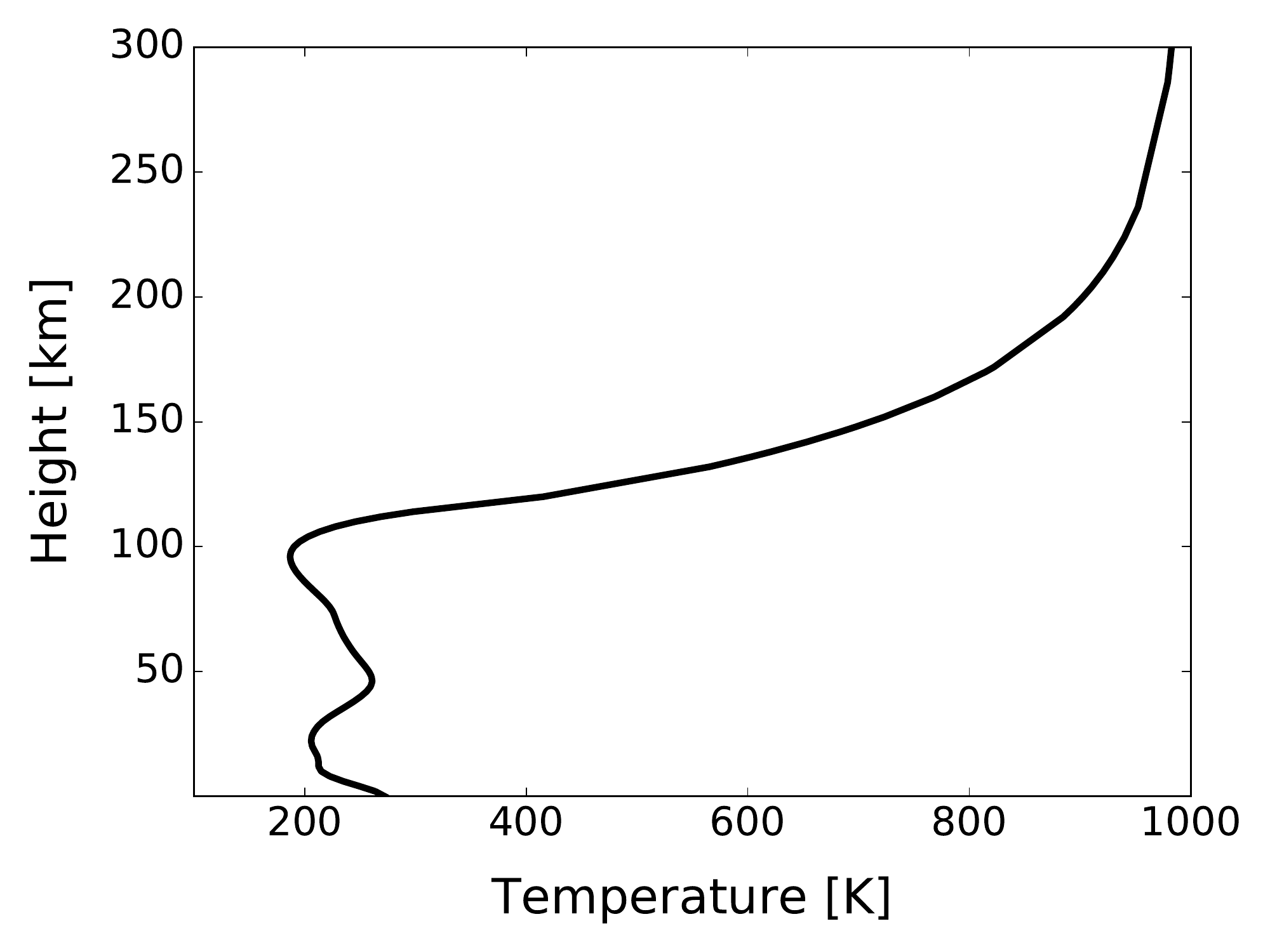}
    \caption{Temperature profile of Earth \citep{Hedin1987,Hedin1991}.}
\label{fig:Tprofile}
\end{figure}

To study the impact of lightning on the atmospheric chemistry, we firstly assume that the pressure and temperature profiles of both Contemporary and Early Earth are identical, consistent with \citet{Rimmer2016}. The temperature profile of the Earth atmosphere is adapted from \cite{Hedin1987, Hedin1991} as shown in Figure \ref{fig:Tprofile}.

The chemical composition of present-day Earth is oxidising, dominated by nitrogen (80\%) and oxygen (20\%) with traces of other elements. The chemical abundances at 0 km altitude (1 atm) are summarised in Table \ref{earth-composition} and are chosen according to \citet{Seinfeld2016}.

\begin{table}
\centering
\caption{Chemistry at the surface of the Contemporary and Early Earth used in our models.}
\label{earth-composition}
\begin{tabular}{lll}
Chemical species & Contemporary Earth & Early Earth \\ \hline
N$_2$   & 0.8                   &   0.8 \\
O$_2$   & 0.2                   &   0.0  \\
H$_2$O  & 0.01            &  0.01   \\
CO$_2$  & 3.5$\cdot10^{-4}$     &  0.1   \\
CH$_4$  & 2.0$\cdot10^{-6}$     &  0.0   \\
H$_2$   & $1.0\cdot10^{-6}$     &  $1.0\cdot10^{-3}$    \\
CO  & $6.0\cdot10^{-7}$     &  $5.0\cdot10^{-5}$   \\
N$_2$O  & $6.0\cdot10^{-7}$     &  0.0   \\
O$_3$   & $2.0\cdot10^{-8}$     &  0.0   \\
HNO$_3$ & $1.0\cdot10^{-10}$    &  0.0   \\
NO$_2$  & $6.0\cdot10^{-11}$    &  0.0   \\
NO      & $2.5\cdot10^{-11}$    &  0.0   \\
HO$_2$  & $7.233\cdot10^{-12}$  &  0.0   \\
HO      & $7.2333\cdot10^{-14}$ &  0.0  \\ \hline
\end{tabular}
\end{table}

For Early Earth, we adapt the the atmospheric composition introduced by \cite{Kasting1993}. His weakly reducing atmosphere is the best simultaneous explanation of the observed hydrogen fractionation, $^{22}$Ne/$^{20}$Ne and xenon isotope ratios \citep{Hunten1973,Ozima1980,Zahnle1990a,Zahnle1990b}, and is the atmosphere that is most consistent with the best understood atmospheric escape rates of H$_2$ \citep{Lammer2008}. According to \cite{Kasting1993}, the atmosphere of Early Earth is assumed to consist mostly of 80\% of \ce{N_2} and 10\% of \ce{CO_2}, with traces of \ce{H_2O}, \ce{H_2}, and \ce{CO} at 0 km altitude; see Table \ref{earth-composition}.

The cooling function for both atmospheres is estimated as described in Appendix \ref{appendix:app2} for the main chemical constituents -- \ce{N_2} and \ce{O2} for Contemporary Earth, and \ce{N_2} and \ce{CO_2} for Early Earth.

\section{Results}
We use the model discussed earlier to study the practical upper limit of the impact of lightning on present-day and Early Earth atmospheres. The results provide an estimated impact of global super-intense lightning storms on exoplanets similar to Earth and orbiting Sun-like stars. The model also allows us to predict the results of balloon experiments within lightning storms on the Contemporary Earth. In addition, it can be used as a tool for estimating the chemical impact of lightning on the Early Earth. 

\subsection{Contemporary Earth} \label{modern}

We first turn our attention to the \ce{NO_x} production during lightning in order to validate the chemical output of the code. At high temperatures nitric oxide is formed via Zel'dovich reactions \ref{eq:zeld11} and \ref{eq:zeld22}.
\begin{align}
\textup{O}_2+\textup{N} & \rightarrow \textup{NO}+\textup{N} \label{eq:zeld11} \\
\textup{N}_2+\textup{O} & \rightarrow \textup{NO}+\textup{O} \label{eq:zeld22}
\end{align}
STAND contains the reverse reaction of Reaction (\ref{eq:zeld22}):
\begin{gather}\label{eq:reverse}
\textup{NO}+\textup{O} \rightarrow \textup{N}_2+\textup{O} 
\end{gather}
The reverse reaction rates will lead the gas into chemical equilibrium if enough time is given and if no disequilibrium processes (eg. photochemistry) are included. Such reaction rates are not always physically accurate, and in our case lead to significant underestimation for reaction \ref{eq:zeld22} at high temperatures. Thus it was decided to use experimentally obtained coefficients from \cite{Michael1992} at high temperatures to calculate the rate coefficient, $k$ [cm$^{3}$ s$^{-1}$], via equation \ref{eq:2bodyrate} where $T$ [K] is the temperature:
\begin{equation} 
\label{eq:2bodyrate}
k(T)=1.66\cdot10^{-10}\, \mathrm{cm^3 s^{-1}} \; e^{-3.8\cdot10^4 \, \mathrm{K}/T}.
\end{equation}
This clearly demonstrates the need for experimental studies that would provide the lacking rate coefficients for this and other missing reactions. Our model shows that Zel'dovich reactions take place only in the temperature range $\approx2000-10100\,\textup{K}$. The rate of formation of \ce{NO} reaches up to $\approx10^{20}$ cm$^{-3}$ s$^{-1}$. 

\begin{figure}
  \includegraphics[clip,width=\columnwidth]{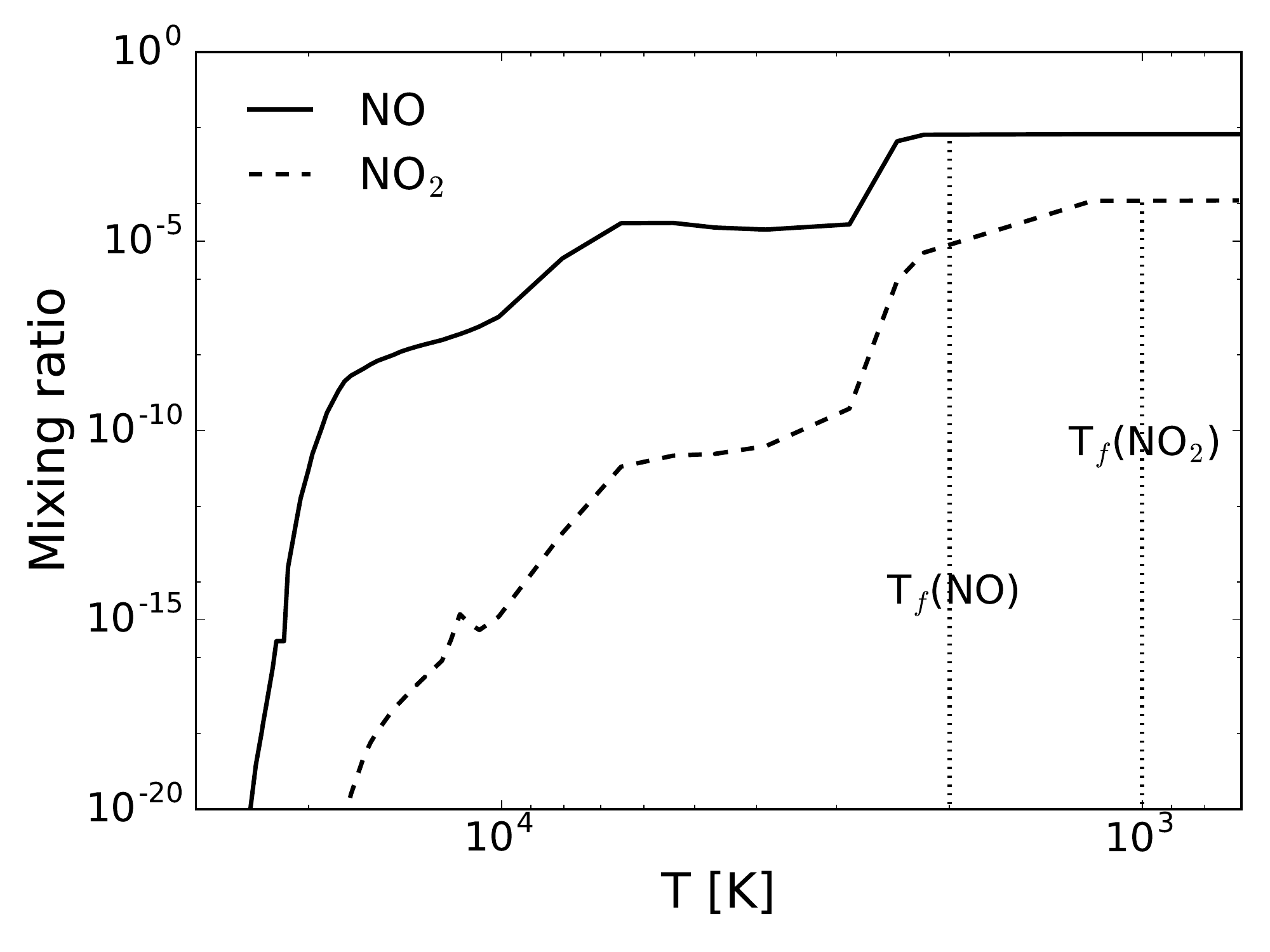}%
\caption{Mixing ratios of \ce{NO} (solid line) and \ce{NO_2} (dashed line) as a function of temperature during lightning for the Contemporary Earth. The mixing ratios reach up to $X(\textup{NO})=6.766\cdot10^{-3}$ and $X(\textup{NO}_2)=1.175\cdot10^{-4}$. The estimates 'freeze-out' temperature are shown in grey dotted line, $T_f(\textup{NO})\approx2000\,K$ and $T_f(\textup{NO}_2)\approx1000\,K$.}
\label{fig:no_no2}
\end{figure}

Another 3-body reaction is consistently observed to produce nitric oxide from the very beginning of the electric discharge, and is:
\begin{gather} \label{eq:3body}
\textup{N}+\textup{O}+\textup{M}\rightarrow \textup{NO}+\textup{M},
\end{gather}
where \ce{M} is any third body.
This reaction disappears only when the heated air returns into
thermodynamic equilibrium and cools down to pre-shocked
temperatures. The importance of this 3-body association (Eq. \ref{eq:3body}) to forming \ce{NO} during a lightning event has not to our knowledge been mentioned anywhere in the literature before now.

Figure \ref{fig:no_no2} demonstrates the change in mixing
ratios\footnote{The mixing ratio of a species X is the number density
  of that species divided by the total number density: $n({\rm
    X})/n_{\rm tot}$.} during the lightning for both \ce{NO} (solid
line) and \ce{NO_2} (dashed line). This allows to estimate the
'freeze-out' temperature, $T_f$, after which almost no
change in the mixing ratios occur (vertical dashed line)
\citep{Navarro-Gonzalez2001}. For \ce{NO}, $T_f(\textup{NO})$ is estimated to be $\approx2000\pm500\,\textup{K}$. The net yield is calculated from our results using Eq. (49) from \cite{Rimmer2016}, and is equal to $P(\textup{NO})=(8.04\pm2.00)\cdot10^{16}\,\textup{molecules}\,\textup{J}^{-1}$. \cite{Borucki1984} predicted the net yield of produced nitric oxide during the lightning discharge to be $(9\pm2)\cdot10^{16}\,\textup{molecule}\,\textup{J}^{-1}$. The
laboratory studies of electric discharge demonstrated the yield of
$(1.5\pm0.5)\cdot10^{17}\,\textup{molecule}\,\textup{J}^{-1}$
\citep{Navarro-Gonzalez2001}. The produced
levels of \ce{NO} are in agreement with both experimental and
observational values.

\begin{figure}
  \includegraphics[clip,width=\columnwidth]{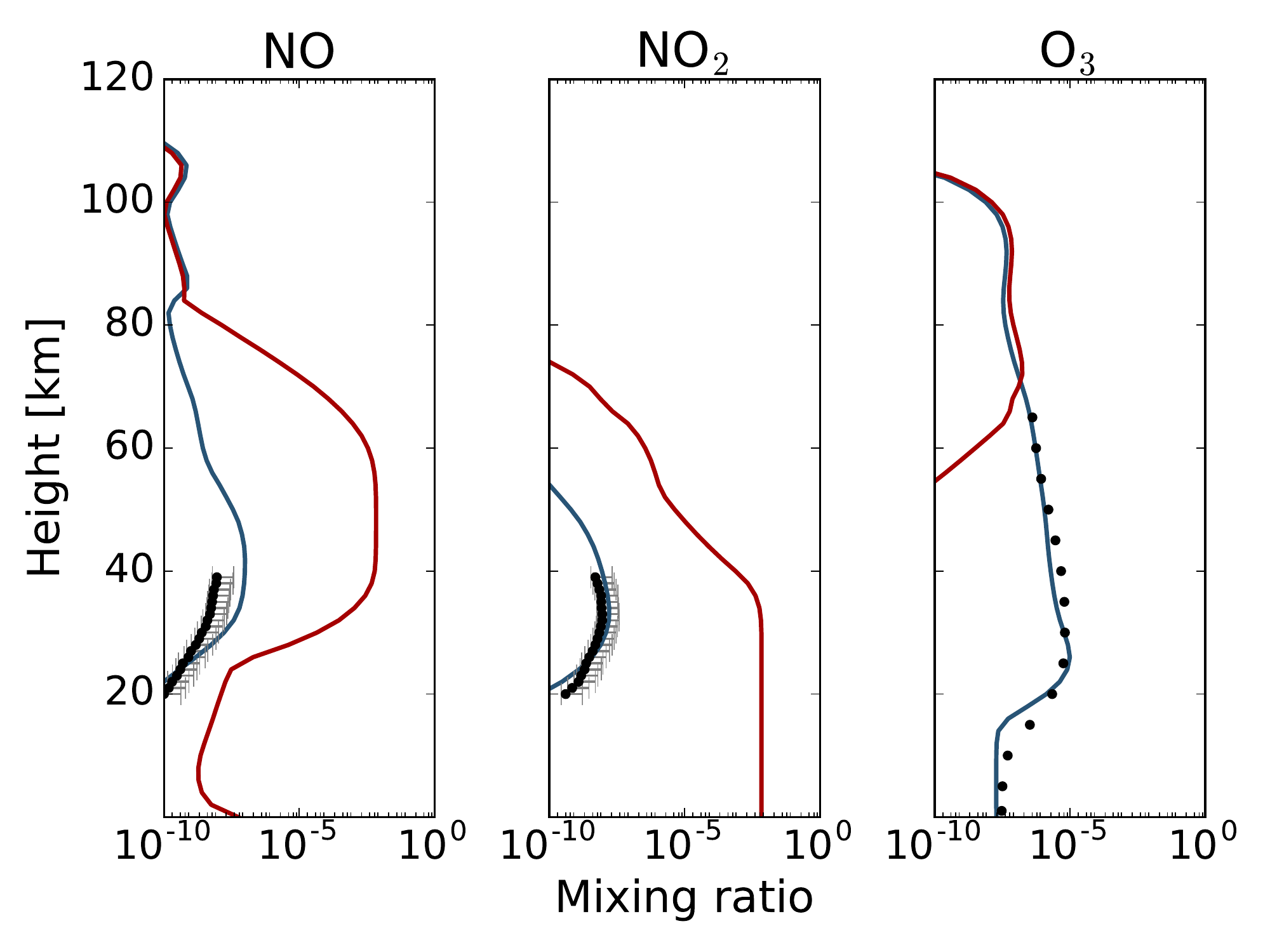}%
\caption{The atmospheric profiles of \ce{NO}, \ce{NO_2}, and \ce{O_3} for the Contemporary Earth.  Red line -- with lightning, blue line -- without, black dots -- balloon observations (\ce{NO} and \ce{NO_2} from \protect\cite{Sen1998}, \ce{O_3} from \protect\cite{Massie1981}). The model demonstrates a good fit with the measurements.}
\label{fig:no_no2_o3}
\end{figure}

The atmospheric profiles for each chemical specie are produced by the model and include the photochemical and diffusion processes. Figure \ref{fig:no_no2_o3} shows the profiles of \ce{NO}, \ce{NO_2}, and \ce{O_3} in the case of lightning (red) and without (blue). The initial increase of \ce{NO} and \ce{NO_2} is a consequence of lightning at 0 km altitude, where the mixing ratios reach $X(\textup{NO})=4.9\cdot10^{-8}$ and $X(\textup{NO}_2)=7.3\cdot10^{-3}$. Nitric oxide is then destroyed to produce \ce{NO_2}, \ce{NO_3},and \ce{N_2O_3}. At 10--60 km altitude, the abundance of \ce{NO} is increasing due to the reverse reactions reaching the maximal value of $X(\textup{NO})=6.5\cdot10^{-3}$. \ce{NO_2} remains constant until 40 km. Higher in the atmosphere, the photochemical reactions destroy both \ce{NO} and \ce{NO_2}. 

The fraction of ozone is visibly reduced by the lightning because most of the oxygen is in the nitric oxide. When the photochemical destruction of \ce{NO} becomes efficient, the mixing ratio of \ce{O_3} reaches its non-lightning value and even slightly exceeds it, reaching $X(\textup{O}_3)=1.9\cdot10^{-7}$ at 75 km. 

The atmospheric profiles in Fig. \ref{fig:no_no2_o3} also include the values obtained during the balloon measurements when no lightning is present in the atmosphere. The simulation results are in a good agreement with the measurements. Thus, we can use the results to predict maximal mixing ratios for balloon measurements taking place from within a lightning storm.

\subsection{Early Earth} \label{Early}

\begin{figure}
  \centering
    \includegraphics[width=\columnwidth]{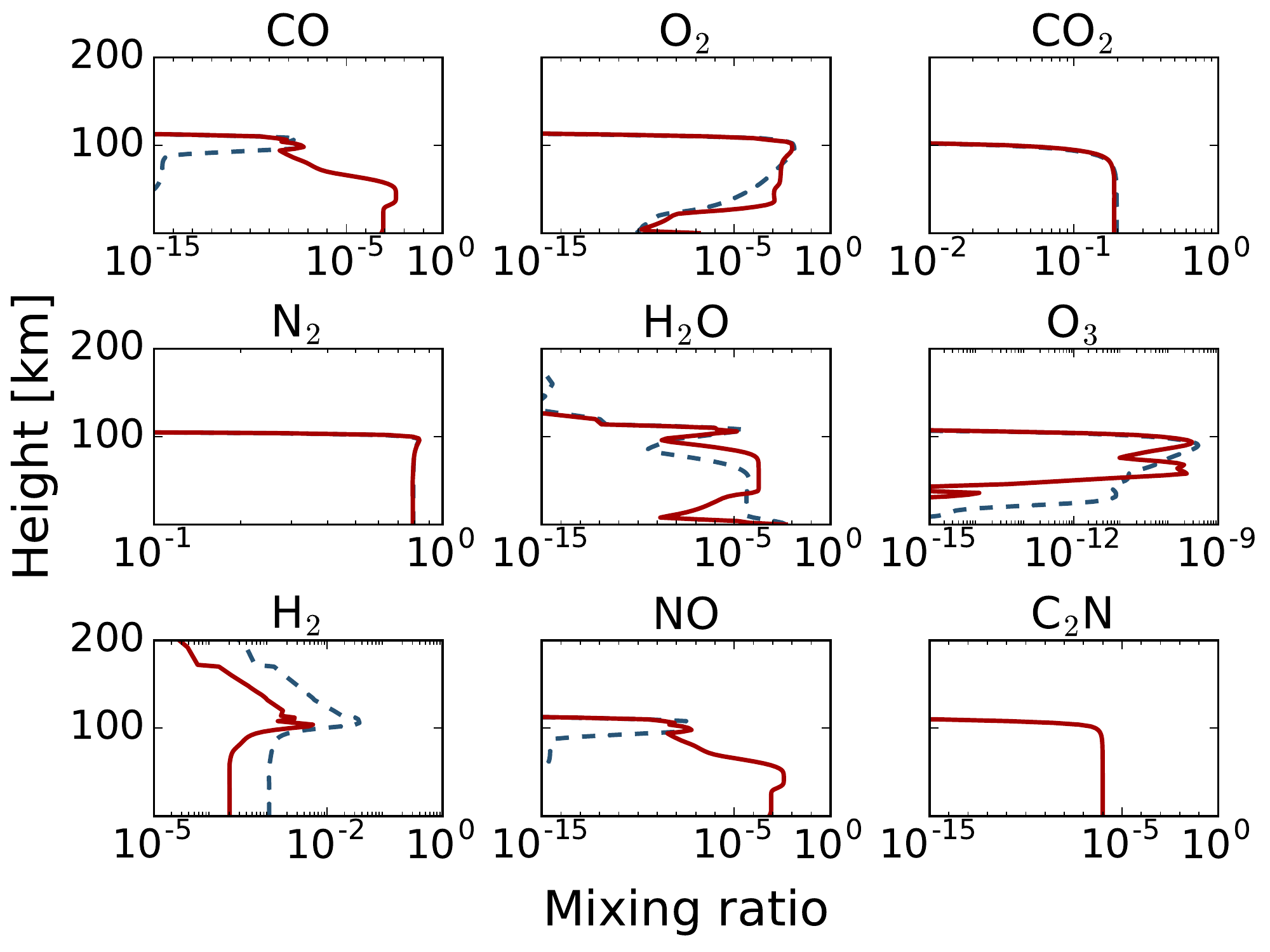}
    \caption{Atmospheric profiles of \ce{CO}, \ce{O_2}, \ce{CO_2}, \ce{N_2}, \ce{H_2O}, \ce{O_3}, \ce{H_2}, \ce{NO}, and \ce{C_2N} for the Early Earth. Blue line -- lightning off, red line -- with lightning on.}
\label{fig:ee_summary}
\end{figure}

We then apply our code to the Early Earth by shocking our parcel at 0 km height within a bulk atmosphere from \citet{Rimmer2016}. We assume an effective lightning flash density of $5.4 \times 10^4$ flashes km$^{-2}$ h$^{-1}$. Figure \ref{fig:ee_summary} shows the variations in atmospheric profiles of the main chemical elements for both with and without lightning.

The abundance of \ce{N_2} and \ce{CO_2} decrease by a very small amount during the lightning discharge. The mixing ratio of \ce{CO} increases initially to $10^{-2}$  and remains constant until gets dissociated by photochemistry. The net amount of \ce{H_2} is decreased by the presence of lightning. This correlates with the production of H-containing molecules during electric discharge, such as \ce{HNO}, \ce{NH_2OH}.

Similar to the present-day Earth case, the simulation demonstrates very efficient formation of nitric oxide during the lightning event. The maximal mixing ratio during electric discharge reaches up to $X(\textup{NO})=1.33\cdot10^{-3}$. The ``freeze-out'' temperature is approximated to be $2190\pm300$ K. Thus, using Eq. (49) from \citet{Rimmer2016}, the estimated yield is $P(\textup{NO})\approx 1.42\pm0.19\cdot10^{16}\,\textup{J}^{-1}$. \cite{Kasting1981} estimated the production efficiency during the lightning event in the Early Earth as  $(0.27-1.1)\cdot10^{16}\,\textup{molecule}\,\textup{J}^{-1}$, assuming the ``freeze-out'' temperature to equal 3500 K. The difference in ``freeze-out'' temperatures for Contemporary and Early Earth atmospheres arise due to the different cooling rates appropriate for the bulk composition of Early Earth. Such a large yield makes both \ce{NO} and \ce{NO_2} possible candidate for the lightning tracer on any Earth-like planet around any star. However, stars with different XUV field will destroy some of these species more or less rapidly.

The computational model shows that different reactions are responsible for \ce{NO} formation compared to the Contemporary Earth atmosphere. Zel'dovich reactions are present, however, reaction \ref{eq:zeld2_ee} occurs only in the narrow temperature range from $\approx 3000 - 2400$ K. This is explained by the overall reduced amount of oxygen present in Early Earth atmosphere. The 3-body reaction is preset from the beginning of the lightning and disappears at $ \approx 2400$ K. At temperatures lower than 2000 K, nitric oxide is formed due to the dissociation of more complex H-rich molecules produced by the lightning, such as \ce{NHO}, \ce{NH_2O}, and \ce{HNO_2} (reactions \ref{eq:hno_ee}, \ref{eq:nh2o_ee}, and \ref{eq:hno2_ee}). 
\begin{gather} \label{eq:zeld1_ee}
\textup{N}_2 + \textup{O} \rightarrow \textup{NO} + \textup{O} \\ \label{eq:zeld2_ee}
\textup{N}+\textup{O}_2 \rightarrow \textup{NO}+\textup{N} \\  \label{eq:3body_ee}
\textup{N}+\textup{O}+\textup{M} \rightarrow \textup{NO} + \textup{M} \\ \label{eq:hno_ee}
\textup{H} + \textup{HNO} \rightarrow \textup{H}_2 + \textup{NO} \\ \label{eq:nh2o_ee}
\textup{NH}_2\textup{O} \rightarrow \textup{H}_2 + \textup{NO} \\ \label{eq:hno2_ee}
\textup{HNO}_2 \rightarrow \textup{HO}+\textup{NO}
\end{gather}
The noticeable deviation from the non-lightning case is also observed in the profile of \ce{C_2N}. The mixing ratio reaches $1.1\cdot10^{-6}$ during the lightning and remains constant up until 100 km. In reality, the destruction of \ce{C_2N} might occur at much lower altitudes in the atmosphere, since this molecule has not been studied in detail experimentally. However, there have been extensive theoretical studies into its reactions with \ce{H_2O}, \ce{CH_4}, \ce{NH_3}, \ce{C_2H_2} \citep{Wang2006} and \ce{H_2S} \citep{Dong2010}. These theoretical studies suggest that reactions between \ce{C_2N} and these species proceed without barriers, but the branching ratios and rate coefficients remain undetermined. There has been some yet unpublished work involving the reaction of \ce{C_2N} with \ce{CO_2}, which is expected to encounter a moderate barrier, and with \ce{NO_2}, which may be proceed efficiently (J. Wang, private communication). Because the rate constants for these reactions remain undetermined, STAND at present does not include these destruction pathways for \ce{C_2N}. We are hopeful that future work will be performed to fix the branching ratios and allow us to estimate reliable rate constants for these destruction pathways, in order to determine both the stability of \ce{C_2N} within the deep atmosphere, as well as the chemical fate of its products. The 'freeze-out' temperature for \ce{C_2N}, sans the destruction pathways, is estimated to be around 4000 K. The formation path from a fully ionised gas is determined shown in Table \ref{cnc}. The reaction in bold corresponds to the rate-limiting step which defines the timescales for the whole reaction chain. We have found no literature relating cyanomethylidyne (\ce{C_2N}) either to lightning or to atmospheric chemistry, although \citet{Wang2006}, among other publications, propose that \ce{C_2N} would plausibly be present in detectable quantities within interstellar clouds and disks. Study of the reaction kinetics of \ce{C_2N} is important for all hydrogen-poor atmospheres where dissociation is important. It doesn't matter whether the dissociation is caused by lightning or geochemistry or biochemistry. 

\begin{table}
\centering
\caption{Balanced path for \ce{C_2N} formation during the lightning. The reaction in bold corresponds to the rate-limiting step.}
\label{cnc}
\begin{tabular}{l}
\ce{2(C^+ + e^- -> C)} \\
\ce{2(N^+ + e^- -> N)} \\
\ce{C + N -> CN} \\
\ce{CN + CN -> NCCN} \\
\textbf{\ce{C^+ + NCCN -> C_2N^+ + CN}}\\
\ce{C_2N^+ + e^- -> C_2 + N} \\
\ce{C_2 + N -> C_2N}  \\ \hline
\ce{5e^- + 3C^+ + 2N^+ -> C_2N + CN}
\end{tabular}
\end{table}

Lightning causes the molecular oxygen to decrease initially since oxygen atoms are more likely to end up in the nitric oxide. Higher in the atmosphere, the fraction of \ce{O_2} increases due to the dissociation of \ce{NO_x} molecules:
\begin{gather}
\textup{NO}_3  \rightarrow \textup{NO} + \textup{O}_2 \\
\textup{O} +\textup{NO}_2 \leftrightarrow  \textup{NO} + \textup{O}_2  \\
\textup{NO}_2 + \textup{NO}_3 \leftrightarrow  \textup{NO}_2 +\textup{NO} + \textup{O}_2
\end{gather}
In the non-lightning case, oxygen forms from hydrogen and carbon containing reactions. These reactions are not observed when the lightning is present. Only reactions involving \ce{NO_x} produce substantial amount up to 100 km. At 24 km, a rapid decrease in the mixing ratio of \ce{O_2} is caused by the interactions involving nitrosyl hydride (\ce{HNO}). \ce{HNO} is formed from the following reactions involving \ce{NH_2O}, and destroys molecular oxygen in the following way:
\begin{gather}
\textup{NH}_2\textup{O} + \textup{HO} \rightarrow \textup{HNO} + \textup{H}_2 \label{eq:nh2o1} \\
\textup{NH}_2\textup{O} + \textup{CHO} \rightarrow \textup{HNO} + \textup{CH}_2\textup{O} \label{eq:nh2o2} \\
\textup{HNO} +\textup{O}_2 \rightarrow  \textup{NO} + \textup{HO}_2 \label{eq:nho}.
\end{gather}
The \ce{NH_2O} is a highly unstable, transitional species of known importance in many chemical kinetics pathways, and is efficiently formed by lightning. It quickly reacts away to nitrosyl hydride (\ce{HNO}) which then is destroyed via oxidation, resulting in a significant decrease of molecular oxygen.

\section{Transmission spectroscopy with Lightning} \label{spectra}
Synthetic transmission spectra in the range of 0.5 - 10\,$\mu$m were computed for an an Earth like planet orbiting a Sun-like star. Four models were computed for the ``Early Earth'' and ``Contemporary Earth'' scenarios, with and without lighting. The 1D radiative transfer forward model of the Tau-REx atmospheric retrieval framework \citep{Waldmann2015,Waldmann2015b}, based on the Tau code by \citet{Hollis2013}, was adapted to compute the transmission spectrum given the variable temperature-pressure profiles and altitude dependent mixing ratios by the STAND network. The transmission spectra are given in terms of the ratio of the radius of the planet, $R_p$ to the radius of the star $R_*$ squared, or $R_p^2/\R_*^2$, and scaled by $10^{-5}$ so that the features can be seen by eye.

\begin{figure}
  \centering
    \includegraphics[width=\columnwidth]{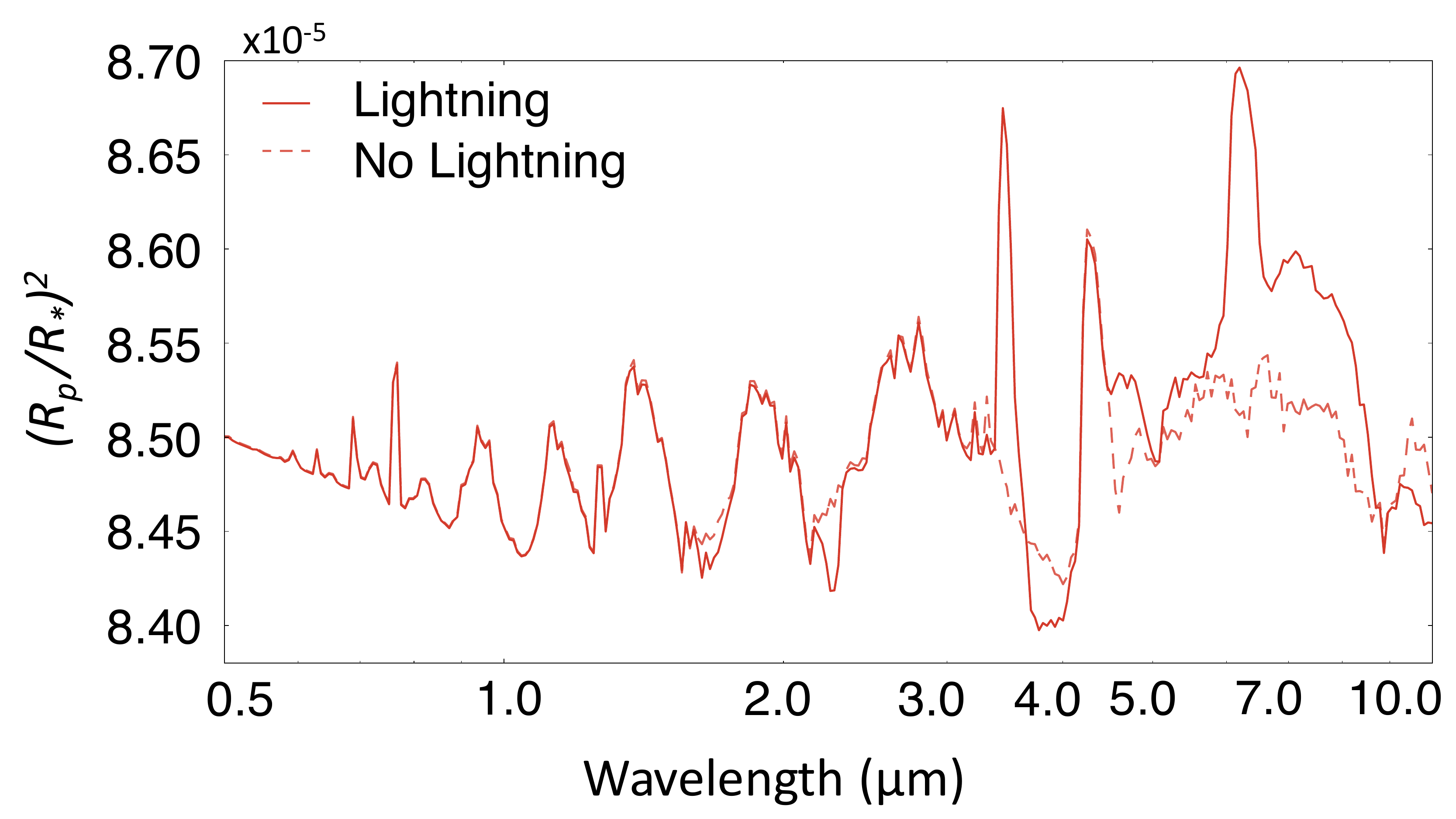}
    \caption{Transmission spectrum of a Contemporary-Earth-like planet 1 AU from a solar-type star, in terms of the transit depth versus the wavelength in microns.}
\label{fig:modern-earth-spec}
\end{figure}

\begin{figure}
  \centering
    \includegraphics[width=\columnwidth]{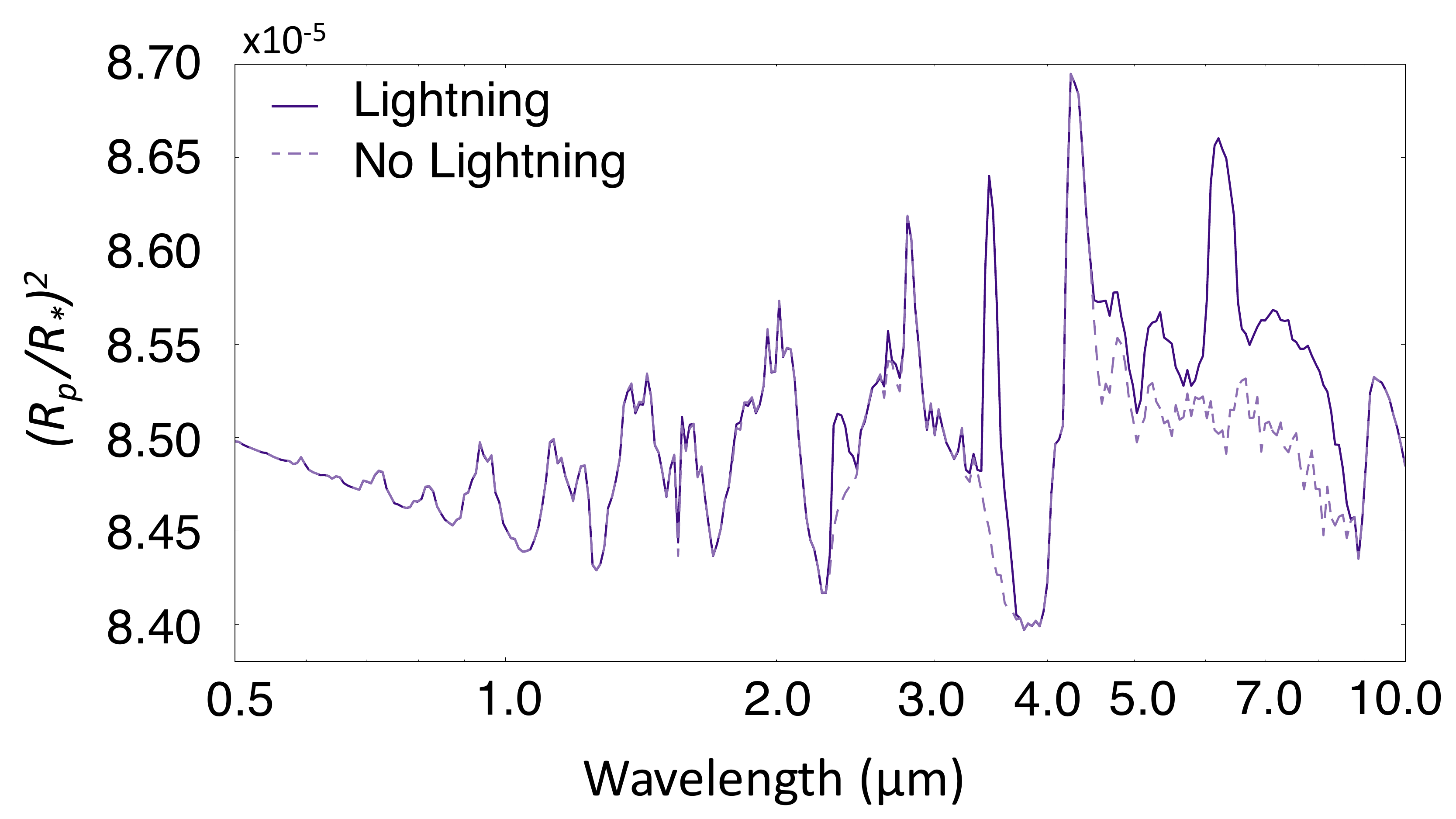}
    \caption{Transmission spectrum of an Early-Earth-like planet 1 AU from a young solar-type star, in terms of the transit depth versus the wavelength in microns.}
\label{fig:Early-earth-spec}
\end{figure}

\begin{figure}
  \centering
    \includegraphics[width=\columnwidth]{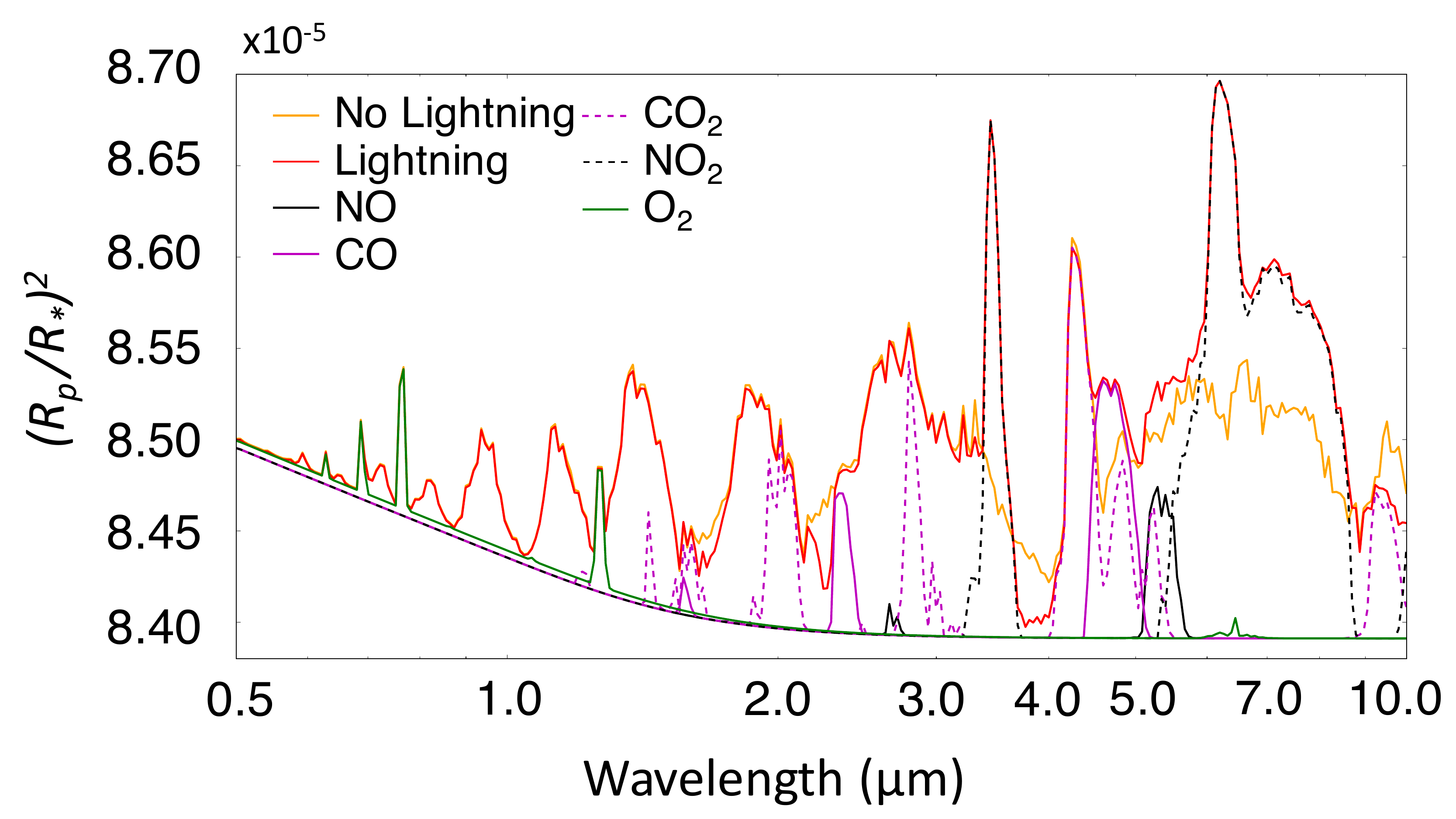}
    \caption{Molecular contributions for the Contemporary Earth transmission spectrum, as a function of wavelength [$\mu$m].}
\label{fig:molecular-opacity}
\end{figure}

Due to the large number of possible opacities of the chemical network, we restricted the computation of the transmission spectra to the most prominent species: O$_2$, O$_3$, NO, NO$_2$, NH$_3$, HCOOH, HCN, H$_2$O, CO$_2$, CO, CH$_4$, C$_2$H$_2$ and C$_2$H$_6$. The mean molecular weight of the atmosphere was calculated using the full chemical network. Temperature and pressure broadened absorption cross-sections were computed at a constant spectral resolution of 7000 and binned to 100 as shown in Figures \ref{fig:modern-earth-spec} and \ref{fig:Early-earth-spec}. Molecular line list opacities were obtained from the ExoMol project \citep{Tennyson2012}, HITRAN \citep{Rothman2009,Rothman2013} and HITEMP \citep{Rothman2010}. Rayleigh scattering and collision induced absorption of H$_2$-H$_2$ and H$_2$-He \citep{Richard2012} were also included. The atmospheres are assumed to be cloud-free. Molecular contributions to the opacity for the ``Modern-Earth'' scenario, both with and without lightning, are shown in Figure \ref{fig:molecular-opacity}.

Figure \ref{fig:molecular-opacity} shows the major contributors of molecular opacity for the Contemporary Earth. The effect lightning has on the spectrum of the Early Earth comes from the same molecular sources as the dominant features. These are \ce{NO} and \ce{NO_2}, \ce{CO} and \ce{CO$_2$} and \ce{O_2}. Although lightning efficiently destroys ozone, it does so in a region where the ozone features are collisionally broadened. The difference in \ce{O_3} opacity changes the transit depth by less than $10^{-8}$. Some differences in
the absorption are present around 4 $\mu$m, affecting the transit depth by a factor of $5\times10^{-7}$, and \ce{CO} absorbs with similar efficiency around 4.5 $\mu$m. The largest effect on the opacity due to nitrogen dioxide (\ce{NO_2}).

Nitrogen dioxide is an efficient absorber at wavelengths of $\sim 3.4$ $\mu$m and $\sim 6.2$ $\mu$m, and increases the transit depth at these wavelengths by at most 2 ppm ($1$ ppm $ = 10^{-6}$). Even with several hours of observation, the James Webb Space Telescope will only be able to resolve changes in transit depth on the order of 10 ppm \citep{Deming2009}, and would need to compete with instrumental systematics and stellar features far larger than the signal itself \citep{Barstow2015}. Detection of the chemical impact of lightning on Earth-like planets, even for the most extreme planet-wide storms \citep[such as Earth-like versions of those discussed by][]{HodosanHAT2016}, will have to wait for the next generation of telescopes, such as the ELT \citep{Gilmozzi2008}. Although the $6.2$ $\mu$m feature will be obscured by the atmosphere, the $\sim 3.4$ $\mu$m feature lies roughly within an atmospheric window, and could be observable from the ground with this kind of future instrumentation. It would then be important to determine what effects reducing the scale of the lightning storm would have on these spectral features, and whether these features would be at all detectable with future instrumentation for more Earth-like thunderstorms.

\section{Conclusions}
We have created the model that can be used to study the impact of lightning for a variety of exoplanetary atmospheres that differ both physically and chemically. We apply this model to an Earth-like rocky exoplanet with both the bulk composition of the Contemporary Earth (\ce{N_2} and \ce{O_2}) and the Early Earth (\ce{N_2} and \ce{CO_2}). We compare our results for the Contemporary Earth, and find that our predictions agree with the experimental and observational yields of \ce{NO} and \ce{NO_2} from lightning. We also make predictions for lightning-induced chemical profiles of \ce{NO}, \ce{NO_2} and \ce{O_3} within thunder clouds.

We show that \ce{NO} is efficiently formed during the lightning via Zel'dovich reactions at temperatures below 10000 K. The studies of Contemporary Earth atmosphere showed, that the considerable contribution to the nitric oxide formation is made by a 3-body reaction \ref{eq:3body}. This reaction is present from the very beginning of the lightning. No information has been found in the literature relating this three-body association to the electric discharge.

For the Early Earth, we find also an enhancement in \ce{NO} and \ce{NO_2}, as well as \ce{CO}. The destruction of \ce{O_3} by lightning is not as important for the Early Earth because comparatively very little \ce{O_3} is predicted within this atmosphere to begin with. We predict also a large production of cyanomethylidyne (\ce{C_2N}), a species which is also predicted to be present within the interstellar medium. \citet{Wang2006} and others have calculated various reaction pathways for \ce{C_2N}, but thus far no reliable rate constants or branching ratios have been published for these reactions. \ce{C_2N} is sufficiently abundant to potentially have an important impact on the atmosphere, either as a spectral signature of lightning, or via the products of its destruction. Further laboratory and theoretical work on this species will be necessary to determine its fate.

Finally, we explored the effect of these species on hypothetical transmission spectra for rocky planets of Earth size with these model atmospheres. Providing an extreme case for the flash density, and therefore chemical yield, we found that, for rocky planets with global and very active lightning storms, the spectrum changes substantially at 3.4 $\mu$m and 6.2 $\mu$m, but these differences are too small to be plausibly detected with JWST, and will have to wait for a future generation of telescopes. Thus, implementation of more physical lightning flash densities will only reduce the already small effect, and will not be relevant for observers unless lightning energetics is very different on other rocky exoplanets than on Earth. The observed spectra will also depend on the composition of clouds which are not included in the model. Incorporation of lightning event rates and clouds can be the next steps for the proper spectra estimation. In the meantime, the tool we have developed for the exoplanet community can be applied to the atmospheres of both hot and cold Jupiters, Brown Dwarfs and mini-Neptunes. If the variability due to lightning is of the same order as the magnitude of the spectral features, as we predict for rocky exoplanets, features of global lightning storms may be observable in these objects.

\section*{Acknowledgements}

AA, PBR and ChH gratefully acknowledge the support of the ERC Starting Grant \#257431. IW, MR, SNY and JT also gratefully acknowledge the support of the STFC (ST/K502406/1), and the ERC projects ExoMol (26719) and ExoLights (617119). PBR thanks John Sutherland and Y.-H. Ding for helpful comments about the chemistry.




\bibliographystyle{mnras}
\bibliography{bibliography} 




\appendix 

\section{Equations of continuity} \label{appendix:app1}

ATHENA solves the equations of hydrodynamics, including a cooling term, and the equation of state \ref{eq:state}. Hydrodynamics equations respect a principle of mass \ref{eq:mass}, momentum \ref{eq:momentum}, and energy \ref{eq:energyideal} conservation, where $\rho$ -- mass density [kg m$^{-3}$], $\mathbf{v}$ -- velocity vector [m/s], $P$ -- pressure [Pa], $E$ -- total energy [J], $\gamma$ -- ratio of heat capacities, and $\Lambda$ -- radiative cooling function [J m$^{-3}$ s$^{-1}$] (see Appendix \label{app2}) \citep{Stone2008}.

\begin{equation}\label{eq:mass}
\frac{\partial \rho }{\partial t}+\bigtriangledown.[\rho \mathbf{v}]=0
\end{equation}
\begin{equation}\label{eq:momentum}
\frac{\partial (\rho \mathbf{ v}) }{\partial t}+\bigtriangledown.[\rho \mathbf{v} \mathbf{v} + P]=0
\end{equation}
\begin{equation}\label{eq:energyideal}
\frac{\partial E }{\partial t}+\bigtriangledown.[(E+P) \mathbf{v}]=-\rho^2\Lambda
\end{equation}
\begin{equation}\label{eq:state}
E=\frac{P}{\gamma -1}+\frac{\rho (\textbf{v.v})}{2}
\end{equation}


\section{Radiative cooling} \label{appendix:app2}

The temperature dependence of lightning shocks is predominately due to radiative cooling. If not for radiative cooling, the time-scale for the temperature to decrease from $\sim 30000$ K to $\sim 10000$ K would be on the order of seconds, rather than microseconds. Predicting accurate cooling rates ab initio would depend on detailed microphysics for various compositions at temperatures and densities not yet investigated, and would require a fully coupled and self-consistent lightning chemistry and radiative hydrodynamics model, which is beyond present computational capabilities. 

 For this paper, we instead take a phenomenological approach to the radiative cooling function, appropriate for a high density high temperature plasma resulting from the lightning shock. We take a low-density approximation as our leading term and then add higher order correction terms to account for the high plasma density that exists within a lightning shock.

As explained above (Section \ref{intro}), at the initial temperature of 30000 K for the centre of the lightning shock, we assume an artificial initial state where all molecules have completely dissociated away, and all remaining atoms are ionised. Collisional ionisation dominates, and to second order, is balanced by the total recombination rate, satisfying the conditions for ``coronal equilibrium''. In effect, the leading term is obtained under the assumption that every excitation is collisional and not due to reabsorbing emitted light, and that radiative, rather than collisional, de-excitation dominates. We therefore use the parametrised cooling functions of \cite{Post1977}, which account for free-free emission, emission from (bound-free) radiative recombination and cooling from line (bound-bound) emission.

The second order cooling rates, $\Lambda_1$ [cm$^{3}$ s$^{-1}$] are given for a single species X by the polynomial:
\begin{equation}
\log_{10} \Lambda_2 ({\rm X}) = \sum_{i=0}^{5} A_i({\rm X}) t^{i},
\label{eq:App-crspec}
\end{equation}
where $t = \log_{10} \big[T_e/(1$ keV$)\big]$ and $T_e$ [keV] is the electron temperature. For multiple species, we sum the mixing ratios of that species, remembering that every constituent in the atmosphere is completely dissociated into its atomic form and then ionized, such that the entire gas is comprised of electrons and cations. The volume mixing ratio is represented for cationic species X by $x(X) = n(X)/n_{\rm cat}$ where $n({\rm X})$ [cm$^{-3}$] is the number density of the cation and $n_{\rm cat}$ [cm$^{-3}$] is the sum of all cations in the gas: $n_{\rm cat} + n(e^-) = n_{\rm tot}$. For our purposes, we consider the gas to be comprised of cations from the three atoms C, N and O. Therefore, the Eq.(\ref{eq:App-crspec}) for each species is weighted by its cationic mixing ratio and then summed:
\begin{align}
    \Lambda_2 &= \sum_{i=0}^5 x({\rm C}) A_i({\rm C}) t^i + \sum_{i=0}^5 x({\rm N}) A_i({\rm N}) t^i + \sum_{i=0}^5 x({\rm O}) A_i({\rm O}) t^i, \notag\\
    &= \sum_{i=0}^5 \Big[x({\rm C}) A_i({\rm C}) +  x({\rm N}) A_i({\rm N}) +  x({\rm O}) A_i({\rm O})\Big] t^i.
\end{align}
These values in the brackets can be represented by a single coefficient relevant for the atmosphere in question, $B_i$, such that:
\begin{align}
\log_{10} \Lambda_2 &= \sum_{i=0}^{5} B_i t^{i}, \notag\\
B_i &= x({\rm C}) A_i({\rm C}) +  x({\rm N}) A_i({\rm N}) +  x({\rm O}) A_i({\rm O}). \label{eq:App-crtab}
\end{align}

The values we use for $B_i$ can be found in \ref{tab:App-coeff}.
    \begin{table*}
    \caption{$A_i$ Coefficients and Mixing Ratios for the Cooling Rate, Eq.(\protect\ref{eq:App-crtab})}
    \label{tab:App-coeff}
    \begin{tabular}{rcccccccc}
    \hline
     X & $\bm{A_0}$ & $\bm{A_1}$ &$\bm{A_2}$ &$\bm{A_3}$ &$\bm{A_4}$ &$\bm{A_5}$&$x_1({\rm X})^*$&$x_2({\rm X})^*$ \\
     \hline
         C & 1970. & 4570. & 4160. & 1870. & 417. & 37.0 & 0.0 & 0.09 \\
         N & -197. & -243. & -74.5 & 31.3 & 21.7 & 3.30 & 0.8 & 0.18 \\
         O & 652. & 1840. & 1980. & 1060. & 280. & 29.3 & 0.2 & 0.73 \\
         \hline
    \end{tabular}
    \begin{tablenotes}
    \small
    \item $^*x_1$ are the cation mixing ratios for the Contemporary Earth and $x_2$ are the cation mixing ratios for the Early Earth.
    \end{tablenotes}
    \end{table*}

At low enough densities, this leading order term, $\Lambda_2$, dominates. Above a certain critical density, $n_c$ [cm$^{-3}$], collisional cooling becomes important as
well as re-absorption of emitted energy. The medium ceases to become transparent to its own radiation. We simply take the critical cooling rate suggested by the upper limit of \cite{Post1977}, $n_c = 10^{16}$ cm$^{-3}$. By analogy to many-body chemical reactions, we modify the cooling term as follows:
\begin{equation}
\Lambda = \dfrac{\Lambda_2}{1 + \sum_{k=1}^{k_{\rm max}} \Big(\frac{n}{n_c}\Big)^{k}}.
\label{eq:App-Cooling}
\end{equation}
The number of terms to be summed, the value of $k_{\rm max}$, would be set by the detailed microphysics. This would take the form of temperature and pressure-dependent higher order cooling rates. Additionally, as the gas cools, eventually atoms react to form complex molecules. These molecules will have different cooling rates than the atoms. 

\begin{figure}
  \centering
    \includegraphics[width=\columnwidth]{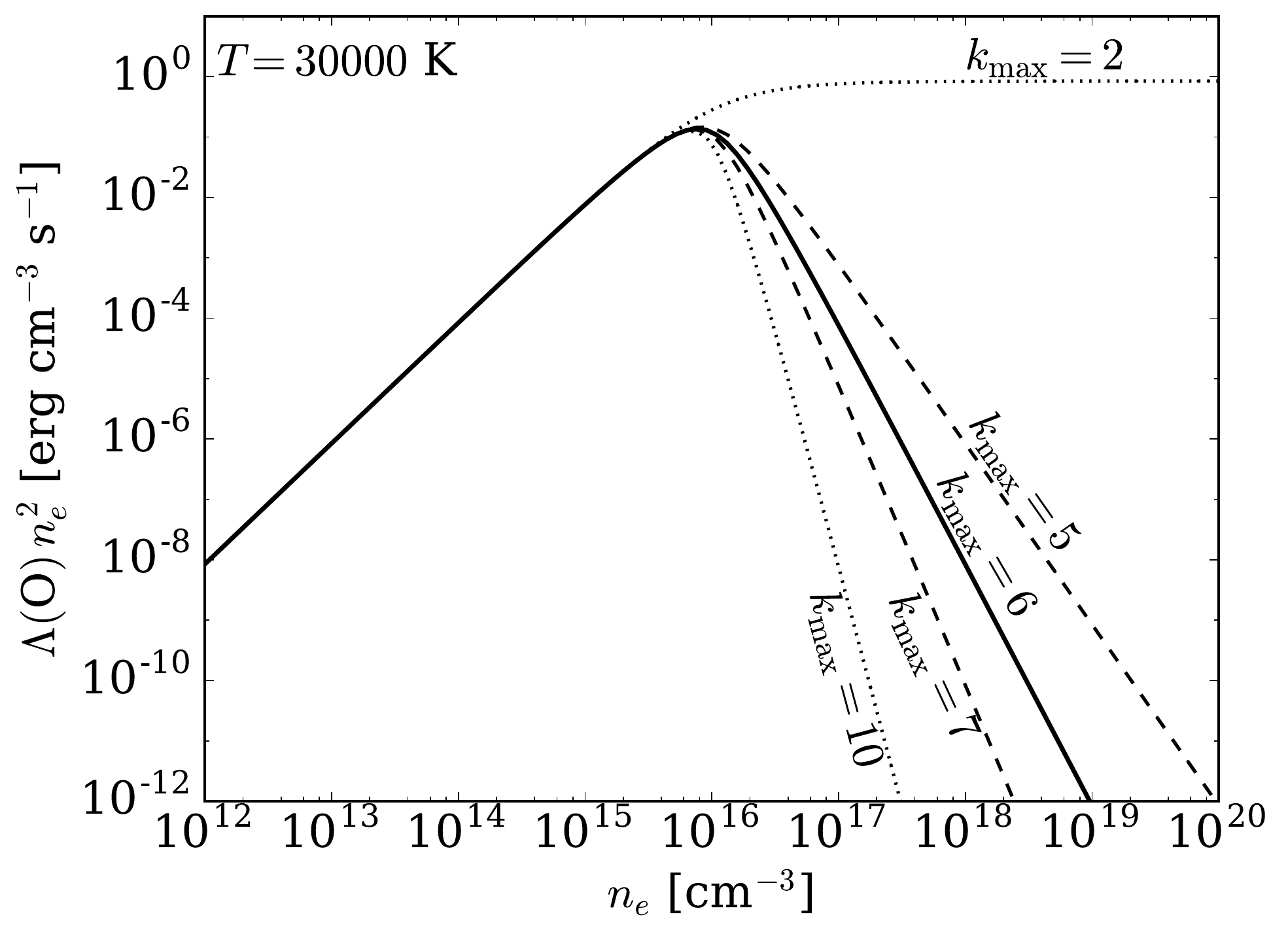}
    \caption{The total cooling rate, $\Lambda \, n_e^2$ [erg cm$^{-3}$ s$^{-1}$] for Oxygen, as a function of electron density $n_e$ [cm$^{-3}$], from Eq. (\ref{eq:App-Cooling}), with different values of $k_{\rm max}$ ranging from 2 to 10 with $k_{\rm max} = 6$, the value used in the rest of this paper, represented with a solid line. Without these corrections, the cooling rate will increase when $n_e > 10^{16}$ cm$^{-3}$ at the same slope as when $n_e < 10^{16}$ cm$^{-3}$.}
\label{fig:cooling}
\end{figure}

Recalling that our method is phenomenological, we set $k_{\rm max}$ to whatever value reproduces the temperature observations from \cite{Orville1968b}. Testing values from 1 to 10, we found that $k_{\rm max} = 6$ best reproduces the observed temperature dependence. Cooling rates for different values of $k_{\rm max}$ are shown in Figure X. The quality of the fit with $k_{\rm max} = 6$ is shown in Figure 2. The dependency of the cooling rate on density in Eq. (\ref{eq:App-Cooling}) agrees with the density dependency found in rigorous microphysical investigations into cooling rates for the range of investigated densities overlap with the proposed cooling rate \citep[see, e.g.][]{Woitke1996}. Because of the physical density dependency of our cooling rate and the reasonable agreement between our cooling rate and chemical and physical observation of lightning on Earth, we have good reason to think that we are representing the cooling from a lightning shock with sufficient accuracy for our work. 


\bsp	
\label{lastpage}
\end{document}